\documentstyle[12pt,aasms4]{article}
\def\apj #1 #2 #3 {#1, ApJ, #2, #3}
\begin{document}
\title{\bf Production of $^2$H, $^3$He, and $^7$Li from Interactions Between 
Jets and Clouds}

\author{M. Famiano\altaffilmark{1}, J. Vandegriff\altaffilmark{1}, R.N. 
Boyd\altaffilmark{1,2}}
\authoremail{famiano@mps.ohio-state.edu}
\authoremail{jonv@mps.ohio-state.edu}
\authoremail{boyd@mps.ohio-state.edu}
\author{T. Kajino\altaffilmark{3-5}}
\authoremail{kajino@th.nao.ac.jp}
\author{P. Osmer\altaffilmark{1}}
\authoremail{posmer@astronomy.mps.ohio-state.edu}

\altaffiltext{1}{Department of Physics, Ohio State University, Smith Lab, 
174 W. 18$^{th}$ Ave., Columbus, Ohio 43210}
\altaffiltext{2}{Department of Astronomy, Ohio State University, McPherson
Laboratory, 140 W. 18$^{th}$ Ave., Columbus, Ohio 43210-1173}
\altaffiltext{3}{National Astronomical Observatory, 2-21-1 Osawa, Mitaka,
Tokyo 181-8588}
\altaffiltext{4}{Department of Astronomy, University of Tokyo, 7-3-1 Hongo,
Bunkyo-ku, Tokyo 113-0033}
\altaffiltext{5}{Department of Astronomical Science, Graduate University for
Advanced Studies, 2-21-1 Osawa, Mitaka, Tokyo 181-8588}

\begin{abstract}
The interactions between jets of high-energy nuclei and nuclei of the 
surrounding medium are studied. Such interactions could be initiated by 
jets from active galactic nuclei interacting with surrounding cool clouds.
The resulting nuclear interactions are found to produce copious amounts of 
$^2$H and $^3$He from the $^4$He nuclei. These results suggest that jets 
of particles from quasars could have produced anomalously high abundances in 
surrounding clouds of some of the nuclides usually thought to characterize big 
bang nucleosynthesis, specifically, the $^2$H seen in absorption spectra.
\end{abstract}


The abundances of the nuclides $^2$H and $^3$He have long provided signatures
of big bang nucleosynthesis (\cite{hata95}, \cite{copi95}). In recent years, 
absorption lines from clouds along the line of sight from a quasar to earth 
have been used to determine the primordial $^2$H abundance (\cite{burles98}, 
\cite{burles99}, \cite{tytler99}, \cite{molaro99}, \cite{kirkman00}). The 
values obtained from these studies were found to be consistent with the 
traditional value. However, the same technique has also resulted in some 
``primordial" $^2$H abundance values (\cite{songaila94}, \cite{carswell94}, 
\cite{rugers96}) that were up to an order of magnitude larger. Subsequent 
reanalysis 
of these systems has suggested that the high $^2$H abundances do not represent 
the primordial values, (\cite{burles99a}), but the issue is not completely 
resolved (\cite{tosi98}).  The question of whether of not the high $^2$H 
abundance is primordial has been addressed from the standpoint of primordial 
inhomogenieties in the baryon-to-photon ratio, with the conclusion being that
if such inhomogenieties are not responsible for the observed $^2$H abundance,
then processes occurring after the initial primordial nucleosynthesis
must be responsible.  Two such processes are stochastic and anomalous
chemical evolution in Lyman limit systems (\cite{fuller97}). 
Photoerosion reactions, induced by photons from, e.g., an accreting black 
hole, on atomic nuclei have also been suggested as a possible source for the 
production of light elements (\cite{boyd89}).  Another proposed source of 
processing is the photon flux from accreting black holes that might have been
created from the collapse of an early generation of massive stars formed 
shortly after decoupling (\cite{gnedin92},\cite{gnedin95}).  However, this 
model has difficulty in predicting abundances that match observation 
(\cite{balbes96}). In this Letter we show that spallation production of these 
light nuclides would be the {\it inevitable} result of jets of high-energy 
nuclei hitting nuclei in the surrounding medium contained, e.g., in cool 
clouds, and that 
such interactions could readily produce enough $^2$H in the cloud to explain
the high $^2$H observation. Since both required entities are expected in many
quasars, high $^2$H abundances might be occasionally expected. However, a range
of abundances from primordial to roughly the high values could be produced by
this mechanism.

We have studied the production of $^2$H and $^3$He from interactions between 
jets and clouds, both entities commonly associated with quasars and AGNs. The 
$^2$H and $^3$He would be produced from interactions between $^1$H amd $^4$He 
from a jet interacting with $^1$H and $^4$He (assumed primordial abundances) 
in ambient gas. A localized buildup of $^2$H and $^3$He would result 
from these interactions and, as we show below, could readily lead to $^2$H and 
$^3$H abundances more than an order of magnitude larger than the ``low" 
primordial $^2$H abundance. Such abundances might then produce the observed 
high $^2$H absorption features from very distant quasars, so could provide a 
natural explanation for the origin of the ``high" $^2$H abundance.

We have assumed parameters for the jets and clouds that are typical of those 
seen in AGNs. Typical sizes of clouds around active galactic nuclei are 400 
solar radii, with typical densities of 10$^{11}$ particles cm$^{-3}$ 
(\cite{peterson97}); this corresponds to roughly a 10$^{-6}$ M$_{\odot}$ 
object. For convenience, we have assumed the clouds to be cylinders with equal 
diameters and thicknesses, and with the axis of symmetry in the direction
of the jet. This is about what is required to stop 100 MeV protons, or 400 MeV 
$^4$He nuclei, so it is appropriate to assume that such clouds would stop the 
high-energy nuclei. Since there are thought to be many clouds ($\sim$ 10$^5$) 
about each active galactic nucleus and, presumably, quasar, any jet might well 
process material in many clouds.  For AGN jets, 1 M$_{\odot}$ y$^{-1}$ is a 
plausible value for the jet output (although this doesn't really
matter; the product of the jet intensity that intersects a cloud and the time
during which they interact is what matters), and their breadth would be
expected to exceed the size of a typical cloud by a large factor. Typical jet
radii at the cloud are about 0.01 - 0.1pc (\cite{lobanov97}, 
\cite{peterson97}), 
so the fraction of mass actually incident on the cloud is proportional to the 
fraction of cloud and jet cross sectional areas. Therefore, we have assumed a 
cloud of 10$^{-6}$ M$_{\odot}$ is bombarded by a jet of 100 MeV protons and 
400 MeV $^4$He nuclei at a rate of 10$^{-3}$ M$_{\odot}$ year$^{-1}$ (although
only the total flux matters). 

Production by spallation of $^2$H and $^3$He was examined by means
of a computer code which determines the yields of various spallation reactions
caused by collisions between incident high-energy nuclei and the nuclei at
rest within a stopping medium. As a high-energy projectile enters the stopping
medium and is slowed by electronic energy loss, it may undergo one of several
nuclear reactions in a collision with a target nucleus at rest.  The code
tracks the $^4$He particles through the cloud, calculating the fraction of 
incident particles that undergo each possible reaction that destroys incident 
nuclides and creates other nuclei as reaction products, as well as those that 
come to rest in the stopping medium. The reaction products are also tracked 
through subsequent possible reactions or to their being deposited in the 
stopping medium. The incident material and the reaction products are assumed
to be well mixed in the stopping medium so that, given that the total jet mass
is assumed to be less than 10\% of that of the stopping medium, the cloud
nuclei are not significantly depleted or diluted. 

Spallation cross sections were taken from the literature when available 
(\cite{meyer72}); the reactions included in the code are listed in Table 1. 
However, a paucity of data for many of the reactions 
necessitated calculations of some cross sections. It appears (\cite{dir}) 
that the direct reaction mechanism is primarily responsible for the reactions
\begin{equation}
^4He + p \rightarrow \ ^3H + 2p,
\end{equation}
\begin{equation}
^4He + p \rightarrow \ ^3He + n + p
\end{equation}
\begin{equation}
^4He + p \rightarrow \ ^3He + d.
\end{equation}
Thus the secondary particle energy distributions of the outgoing $^3$He 
and $^2$H particles were calculated by the direct reaction code DWUCK 
(\cite{kunz}). Such calculations require specification of the nuclear optical 
potential parameters. Those for reactions involving protons follow the general
parametrization of Menet et al. (1971) as shown in Perey and Perey (1976). 
While this
parametrization was based on a study of lower energy protons (30 to 60 MeV)
than some of those with which we are concerned,
the fit to higher energy data is quite similar to parameter sets determined
from higher energy data (\cite{perey76}, \cite{schwandt82}). For reactions
involving deuterons, the parametrization of Perey and Perey (1976) was also 
employed,
but with the spin-orbit parameters of Lohr et al. (1974) added on. The use of 
these
parameters gave results that closely matched the experimental results of
Rogers et al. (1969). Potential parameters for $\alpha$-particles followed
Perey and Perey (1976).


\begin{figure}
\label{fig1}
\begin{minipage}{4in}
\epsfxsize=5.5in
\epsfbox{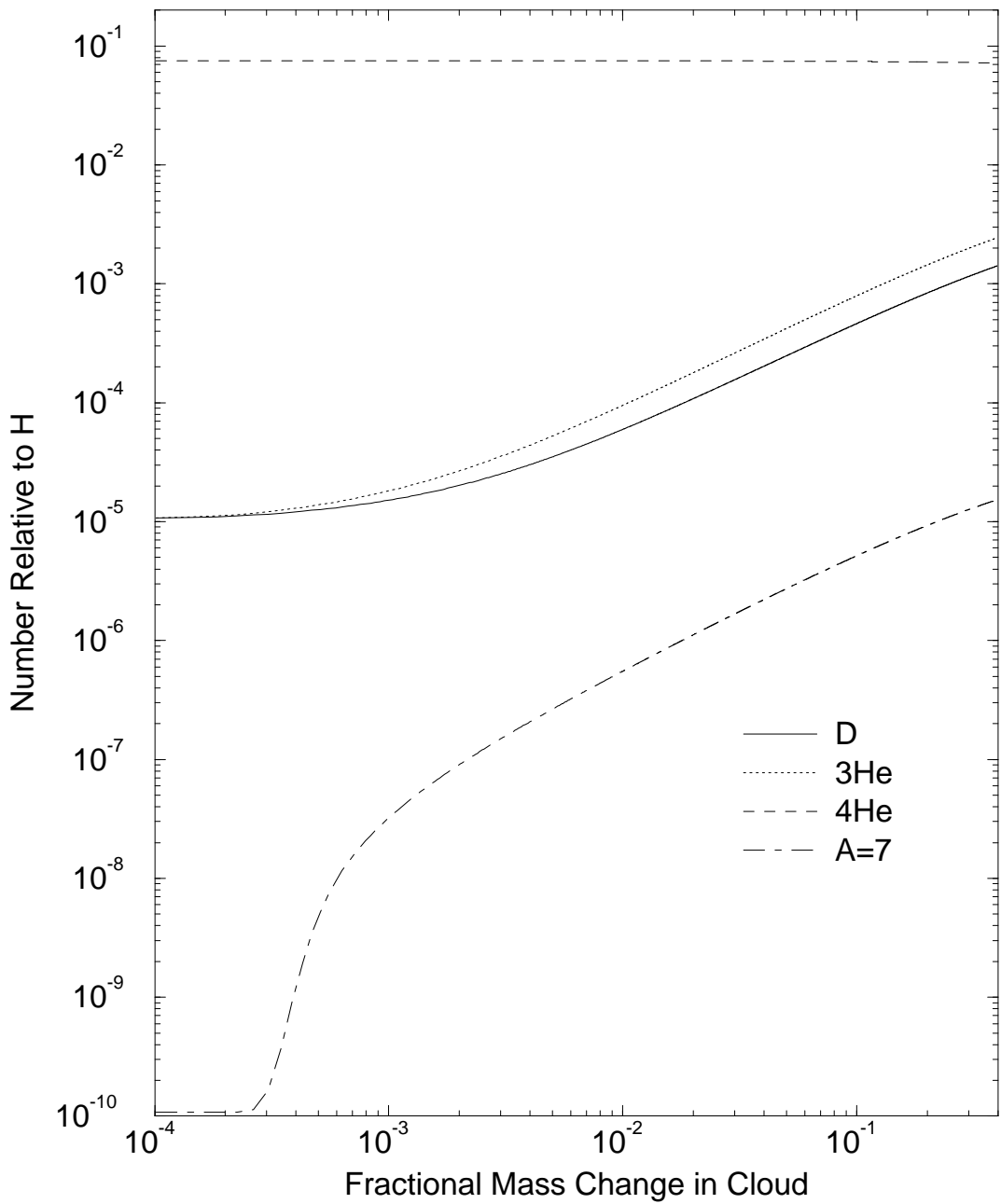}
\end{minipage}
\caption{The final abundances of D, $^3$He, $^3$H, and $^4$He due to the 
spallation of $^4$He into a stopping medium of primordial composition. For 
this figure the initial jet energy is assumed to be 400A MeV.}
\end{figure}

Little information exists for the three-body final state reactions, although
one study (\cite{3body}) indicates that the three-body final states tend to
smear out the structure seen in the energy distributions of the two-body final
state reactions. Some peaking toward the higher energies is suggested by the
one existing data set. Thus we assumed two possibilities as extreme cases: (a) 
one in which the distributions of the reaction products were constant with 
energy, and (b) one in which the distributions were enhanced by a factor of two 
for the particles in the top quarter of the energy distributions. This change 
produced effects in our results only at the 1\% level; the highest energy 
component is small. However, this procedure did allow a test of the sensitivity 
of the results to the shape of these poorly known cross sections. The magnitudes
of the cross sections for the three-body final states are unknown. Thus we 
assumed they scaled with energy as does the  $^1$H($^4$He,$^3$He)d cross 
section; their ratio was then fixed at the one energy at which the 
three-body final state reaction, $^1$H($^4$He,n p)$^3$He, was studied 
(\cite{3body}). Data for the interaction of $^4$He with $^4$He are in 
similarly short supply. Thus we simply assumed that, for such interactions, 
the yields would be three times the values for $^4$He interacting with protons 
at the same center of mass energy, as recommended by Meyer (1972).  

The production of mass 7 elements is included via the reaction 
$^4$He($^4$He,n/p)$^7$Be/Li, and the inverse reaction is also included, with
cross-sections given by Mercer et al. (1997) and Abramovich et al. (1984).
Recent studies on mass 7 nuclei give total reaction cross sections for the 
proton interactions with $^7$Li and $^7$Be (\cite{carlson85}), and excitation 
above the particle emission threshold is obtained by subtracting the inelastic 
scattering cross sections to the first excited states of these nuclei 
(\cite{locard67}). This is appropriate because excitation of states higher 
lying than the first excited state in either $^7$Be
or $^7$Li results in destruction of the nucleus.

Figure \ref{fig1} shows the results of one set of our calculations. The 
results are presented there in a way that allows that graph to be used as a 
universal predictor that depends only on the cross sections used in the 
calculations, i.e., the abundances of the light nuclides are given as a 
function of the fraction of the total mass of the cloud M$_{deposited}$ that 
was added by the jet. The value of M$_{deposited}$ is given by
\begin{equation}
M_{deposited} = (dM/dt)_{jet}  t  f_{overlap}
\end{equation}
where (dM/dt)$_{jet}$ is the rate of mass output within the jet, t is the
time during which the jet interacted with the cloud, and f$_{overlap}$ is the 
fraction of the jet that actually interacts with the 
cloud. As can be seen from Fig. 1, when the mass of the cloud has increased 
by 1\% from the matter added by the jet, the abundances of $^3$He (which
includes that of $^3$H) and 
$^2$H have already increased appreciably, roughly by a factor of four, from 
those normally associated with primordial nucleosynthesis. Furthermore, their 
abundances when 10\% of the mass of the cloud has been added are roughly 
a factor of 30 above their primordial values. Their abundances for this jet 
energy level off gradually as additional mass is added, due to (1) our 
assumption that the density of the cloud remains constant, i.e., the volume of 
the cloud increases with the amount of mass transferred to it by the jet (and 
the cloud is well mixed), and (2) the fact that the fraction of $^4$He nuclei 
that are destroyed depends only on the energy. The values for even larger mass
depositions can be as much as a factor of five higher than those achieved at 
10\% added mass, although amounts of added mass to the cloud in excess of that 
might well destroy the cloud. 

The parameters of the jet and the cloud assumed for this calculation are 
typical of actual systems, as noted above. The cloud mass assumed was 
10$^{-5}$ M$_{\odot}$, the jet output was 1 M$_{\odot}$ y$^{-1}$, and 
f$_{overlap}$ was taken to be 10$^{-3}$. Therefore, the mass deposited into 
the cloud was 0.001 M$_{\odot}$ y$^{-1}$.  In the case of this mass flux, the
cloud mass will increase considerably in a relatively short amount of time.
While this is probably an unrealistically short time scale for a physical 
situation, the relevant parameter is the ratio of the transferred mass to the
total mass of the cloud. The context in which the actual time scale might
matter would be if reactions occurred on time scales that were shorter than or
the order of the $\beta$-decay halflives of the unstable nuclei created: 
$^7$Be and  $^3$H. Assuming, though that those nuclides always decay, rather
than interact again, they will merely contribute to the abundances of $^7$Li
and $^3$He. Thus the abundances of these unstable nuclides are simply summed 
with those of their stable isobars in Figure 1. 

That Figure shows how the $^2$H abundances increases with the amount of
jet mass that interacts with the cloud. Also seen to increase are the
abundances of $^3$He and $^7$Li, in all cases well above their primordial 
values. Thus the jet-cloud interaction 
mechanism is clearly capable of producing large abundances of $^2$H, $^3$He,
and $^7$Li even when the nuclei of the jet and the reaction products from the 
reactions it precipitates are admixed into a considerably larger cloud mass.
 
The energy chosen for the case for which the results are shown in Fig. 1 was 
selected to be well above the thresholds for producing the $^7$Li, $^3$He, 
$^3$H, and $^2$H reaction products. As might be expected, even larger $^2$H,
$^3$He and $^7$Li abundances are observed at higher energies. However, the 
increases  with energy are not large, as the fragile reaction products made 
at the higher  initial energies tend to be destroyed. At 100A MeV 8\% of the 
high-energy $^4$He will be spalled into lighter nuclides, while at 500A MeV 
the fraction increases to 26\%. However, increases in energy do not 
necessarily produce more $^3$He and $^2$H. The production of $^2$H and $^3$He 
peaks at around 500A MeV, as those nuclei, if produced at energies above 500A 
MeV, will essentially all be destroyed by subsequent spallation reactions. We 
have further assumed that the high 
energy particles stop in the medium in which the interactions that produce 
the $^2$H and $^3$He occur. This is clearly not a critical assumption, 
although if the medium were thin enough for the high-energy particles to 
emerge with energies above the reaction thresholds, then the resulting 
$^2$H and $^3$He abundances would decrease accordingly.

Could the clouds that are associated with AGNs or quasars be identified as the 
intergalactic clouds that produce the absorption lines?  A fascinating 
consequence could exist within our own galaxy, the core of which has been
observed to contain much higher deuterium abundances than expected 
(\cite{lubowich00}).  While primordial infall is a suggested possibility, 
the observations are consistent with the possibility that the galactic center 
resulted from jet-cloud interactions in an AGN.  The clouds in which we 
have assumed the processing of $^4$He to $^2$H and $^3$He are close to 
the central engine of the quasar and are at a higher density than those in 
which the absorption occurs. However, the jets that we have assumed to 
interact with the clouds would impart momentum to the clouds, which would be 
loosely bound to the quasar. Although the jet-cloud interaction is complex 
(see, e.g., \cite{wang00}), it does seem plausible that the clouds in which 
the jet-cloud interactions occur could evolve into those in which the $^2$H 
absorption occurs.
Indeed, as noted above, a spread in the values of the $^2$H abundance would be
expected, and this is not inconsistent with that observed (see references in
the introductory paragraph) in distant absorption clouds. Note, though, that 
if the jet-cloud mechanism is found to be important, the values observed for 
$^2$H will provide a lower limit on the primordial deuterium abundance; 
averaging those values would lead to too high a ``primordial" abundance. 

Finally, $^7$Li is predicted to have a very large abundance compared to 
either solar system or primordial abundances. This overabundance would be
consistent with the enhanced Li abundance inferred in the galactic 
center (Lubowich et al. 1998). Note that, if the jet-cloud interaction is the
explanation, this $^7$Li enhancement must accompany the 
enhanced $^2$H abundance; the two are produced concurrently by interactions 
within the same primordial material. Thus the dual enhancement of $^2$H and 
$^7$Li constitutes a test of this model. Mixing subsequent 
to the jet-cloud interaction with material processed in stars would distort 
the ratio of those abundances from those predicted here. Furthermore, the 
actual ratio of the abundances depends on the energy of the particles in the 
jet; the abundance ratio of $^2$H to $^7$Li if the $\alpha$-particles in the 
jet have 200 MeV of energy is 25, 
whereas it is 500 at 1000 MeV. Even with an uncertain amount of mixing and a 
large uncertainty in the energy of the particles in the jet, though, the
qualitative feature of a large dual enhancement would be preserved. 

In summary, the results of calculations describing the interactions between
nuclei in intersecting jets and clouds has shown that large abundances of 
$^2$H, $^3$He, and $^7$Li can be produced therefrom. This might provide an 
explanation for the observed anomalously high ``primordial" $^2$H abundance. 
This mechanism also provides a way for {\it producing} $^2$H, normally thought 
to be only destroyed by galactic chemical evolution. Although prediction of 
the specific enhancements of $^2$H, $^3$He, and $^7$Li that could 
result in clouds is complicated by the several parameters needed to fully 
define the situation, it is clear that the spallation mechanism can produce 
copius quantities of these nuclides.  The possibilities of such 
production might also be expanded to the realm of higher-metallicity regions. 
Many QSO spectra seem to indicate considerable abundances of the CNO elements 
(\cite{hamann99}).  Future work will concentrate on the jet interactions with 
clouds enhanced in these heavier elements.

We note that, if the spallation production of $^2$H and $^3$He is common, it
should be relatively easy to find situations in which these nuclides have been
produced by searching for other absorption lines from quasars. This mechanism
would be expected to produce a wide range of values of the observed $^2$H
abundance, ranging from the true ``primordial'' value to the maximum that 
can be produced by this mechanism, apparently even greater than the highest 
deuterium value yet observed. Such observations could constitute a 
confirmation of the jet-cloud spallation model.

This work was supported in part by NSF grants PHY-9513893 and PHY-9901241. The
authors wish to thank an anonymous referee for several excellent comments and
suggestions.



\clearpage

\begin{table}
\begin{center}
Table 1: Spallation Reactions\\
\begin{tabular}{rlrl}
  \multicolumn{2}{c}{between $^4$He  and $^1$H}     &
                \multicolumn{2}{c}{on secondary projectiles} \\ \hline
        p&\hspace{-4mm}($^4$He,pn)$^3$He &p&\hspace{-4mm}(d,pn)p \\
        p&\hspace{-4mm}($^4$He,2p)$^3$H &p&\hspace{-4mm}($^3$He,dp)p \\
        p&\hspace{-4mm}($^4$He,d)$^3$He &p&\hspace{-4mm}($^3$He,ppn)p \\
        p&\hspace{-4mm}($^4$He,p)2d  &p&\hspace{-4mm}($^3$H,dp)p \\
        p&\hspace{-4mm}($^4$He,2p)dn &p&\hspace{-4mm}($^3$H,pnn)p \\
        p&\hspace{-4mm}($^4$He,2n)3p &$^7$Li&\hspace{-4mm}(p,$^4$He)$^4$He \\
\hline
\end{tabular}
\end{center}
\end{table}

\end{document}